# Vortices in Kekulene Molecules

**L. Menicucci**

Laboratório de Simulação, Departamento de Física, ICEx
Universidade Federal de Minas Gerais, 31720-901 Belo Horizonte, Minas Gerais, Brazil

E-mail: lucamenimeni@gmail.com

**F.C. Sa Barreto**

Departamento de Física e Matemática, Campus Alto Paraopeba, Universidade Federal de São João del Rei, 36420-000, Ouro Branco, Minas Gerais, Brazil and
Emeritus Professor, Laboratório de Simulação, Departamento de Física, ICEx
Universidade Federal de Minas Gerais, 31720-901 Belo Horizonte, Minas Gerais, Brazil

E-mail: fcsabarreto@gmail.com>

**B. V. Costa**

Laboratório de Simulação, Departamento de Física, ICEx
Universidade Federal de Minas Gerais, 31720-901 Belo Horizonte, Minas Gerais, Brazil

E-mail: bvc@fisica.ufmg.br

**Abstract.** Kekulene is an aromatic hydrocarbon with formula $C_{48}H_{24}$ arranged in the shape of a closed super-ring as shown in Fig. 2. It consists of a sublattice with 48 $C$ atoms with spin 5/2 and a 24 hydrogen sublattice with spin 2. In this communication, we use Monte Carlo simulations to determine the magnetic structures present in Kekulene for several temperatures ($T$) and dipole anisotropies ($\delta = D/J$). Our results show that there are two regimes at low temperature separated by a crossover at $2.5 < \delta_{cross} < 3.0$. For $\delta < \delta_{cross}$ the ground state has a unique vortex configuration. In the region $\delta > \delta_{cross}$ arrangements of vortices-antivortices (V-AV) appears. As temperature raises the vortex structure disorders and small oscillations take over. The importance on synthesizing this molecule grounds in the possibility of building real planar structures of sizes at least 10 times smaller than the earlier proposed permalloy nanodots. It is worthy to mention that Kekulene is a planar structure with atomic thickness, which is a great advantage compared with other nanomagnetic structures.

## 1. Introduction

Over the past few years, nanostructures have evoked great interest in several areas ranging from medicine to electronics due to their peculiar mesoscopic properties [1–4] . Of particular importance are magnetic nanostructures candidates for building advanced electronic devices with applications that include magnetic random access memory systems, magnetic sensors, high density magnetic recording media, magnetic head reading and many other [5, 6]. Nano-objects refer to structures that are no more than $100 nm$ in one of its dimensions. To understand their behavior let us remember some properties of magnetic materials. The ground state ($T = 0$) of a finite volume magnetic system has an ordered structure. However, at any non zero temperature the structure is disordered. Specially, the long-range ferromagnetic order in a magnetic device disappears when the energy due to thermal fluctuations becomes comparable to the anisotropic and exchange energy terms [7]. This phenomenon is the well known super-paramagnetic limit imposed on the miniaturization of magnetic devices [8, 9]. The discovery of topologically stable magnetic structures in nano-sized systems has opened a new door to create devices at the nanoscale. Studies on two dimensional materials have shown that the development of topological structures like vortices, skyrmions or hopfions with size comparable to or smaller than the ferromagnetic domain size are good promises to overcome the super paramagnetic limit [10–20]. In some circumstances those objects are long lived meta-stable structures at $T > 0$. A system with only short range exchange interactions, for instance the Heisenberg model, is complectly disordered at any $T > 0$. The introduction of long range forces like dipole or RKKY interactions can produce, together with the exchange interaction, ordered structures at finite temperature in spite of these interactions cannot induce any coherent structure by themselves [7]. A dipole interaction conjugated with exchange interactions is the key ingredient to induce a ground state with a vortex [13]. This fact opens up an entire new area of research looking for nanostructures supporting such non-linear objects. A magnetic vortex (antivortex) is a topological curling structure in which spins on a closed path around the structure core precess by $2\pi$ ($-2\pi$) in the same direction. Direct experimental evidence for its existence was reported by Shinjo et al. [21] using magnetic force microscopy (MFM) to characterize magnetic nanodots of permalloy (NiFe) with a thickness of 50$nm$ and radius between 300$nm$ to 1$\mu m$. The magnetic vortex is characterized by two numbers: the chirality, or vorticity, ($q = \pm 1$) giving the curling direction (clockwise or counterclockwise) and the polarity ($p = \pm 1$) that accounts for the direction of the magnetization at the center of the vortex relative to the vortex plane (Here the $z$ direction.) In Fig. 1 are shown free vortex (Top) and an antivortex (Bottom) in an infinite lattice. Several possible applications of such structures are described in Refs. [22–25]. Four decades ago, in 1978, an aromatic planar molecule named Kekulene was synthesized for the first time by Staab and Diederich [27]. The molecule is an aromatic hydrocarbon with formula $C_{48}H_{24}$ arranged in the shape of a closed super-ring as shown in Fig. 2. It consists of a sublattice with 48 $C$ atoms with spin 5/2 and a 24 hydrogen sublattice with spin 2. Apparently, at that time, there were no interest in exploring the potentialities of this molecule, however, with the increasing interest in nanocarbon structures, due to the growing technological application possibilities, the attention of many groups turned to synthesizing new nanocarbon structures. In 2019 Iago Pozo et al. [26] rediscovered the Kekulene molecule using ultra-high-resolution AFM imaging and provided computational calculations to get information about its molecular structure. The importance on synthesizing this molecule grounds in the possibility of building real planar structures of sizes at least 10 times smaller than the earlier proposed permalloy nanodots. It is worthy to mention that Kekulene is a planar structure with atomic thickness, which is a great advantage compared with other nanomagnetic structures. Because of the spin sizes, Kekulene can be treated in a semiclassical basis. The hamiltonian describing this system can be written as [28, 29]

$$H = -\sum_{\langle i,j \rangle} J_{i,j} \vec{S}_i \vec{S}_j + \sum_i \sum_{j>i} D_{i,j} \left[ \frac{\vec{S}_i \vec{S}_j}{R_{i,j}^3} - 3\frac{(R_{i,j} \cdot \vec{S}_i)(R_{i,j} \cdot \vec{S}_j)}{R_{i,j}^5} \right] \quad (1)$$

where $J_{i,j}$ stands for the exchange interactions between $C-C$, $C-H$ and $H-H$ and $S_i$ represents carbon or hydrogen spins. The indexes $i,j$ label the atomic positions in the structure. The first sum runs over first neighbours and the second over all atoms. $R_{i,j}$ is the distance between sites $i$ and $j$.

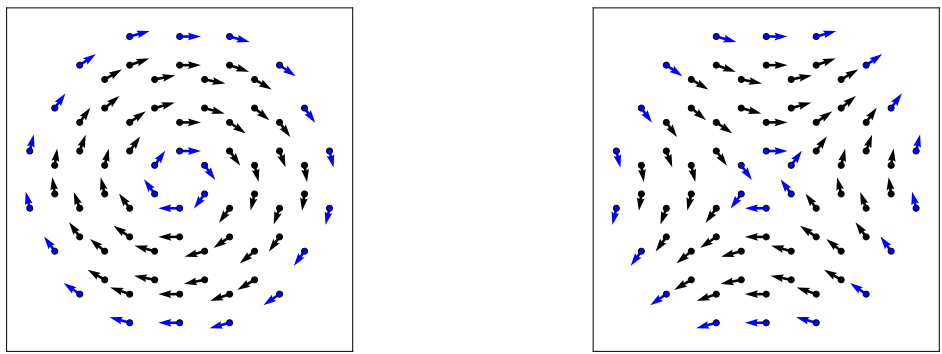

Figure 1: (Color online) In plane structure of a vortex (Left) and an antivortex (Right) in an infinite lattice. Their vorticity is calculated using Eq. 2 as +1 and −1, respectively.

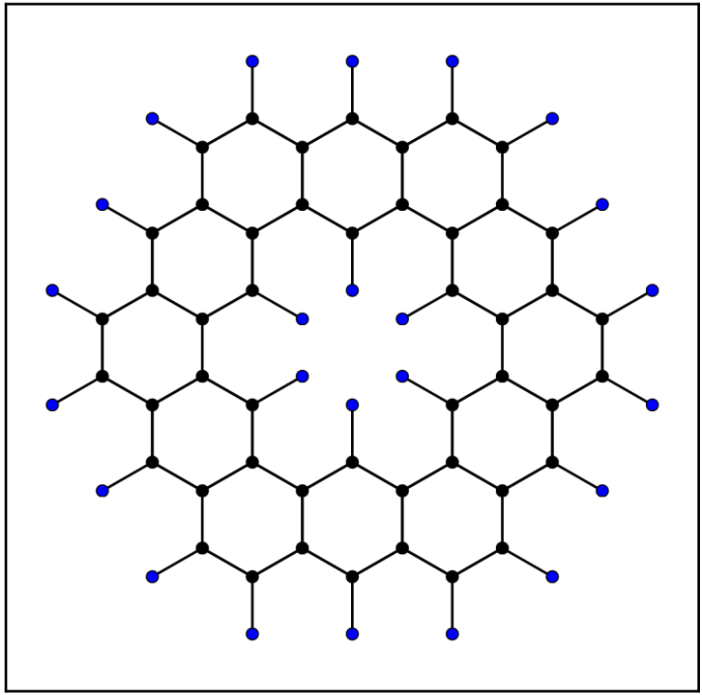

Figure 2: (Color online) Kekulene structure. Black and blue sites are carbon and hydrogen respectively. Distances were calculated by Iago Pozo et al. ([26]).

For simplicity we assume the interaction between atoms is isotropic, so that $J_{ij} = J$ and $D_{ij} = D$ and the distance between first neighbours atoms as isotropic $R = 1$. The second term in Eq. 1 describes the dipole interaction. The exchange interactions in Kekulene are of order of $10^2 kcal/mol \approx 4.8 eV$. We didn't find any information about the dipole interaction. In other systems, like the permalloy, it is at least 10 times smaller than the exchange energy. In this work we will explore the Kekulene properties in a wide range $\delta = D/J$ and temperature. From now on distance and energy will we measured in units of $R$ and $J$ respectively (In other words we take $R =$

1 and $J = 1$). As a matter of simplicity we set the Boltzmann constant to unity $k_B = 1$, so that temperature is measured in units of *energy/$k_B$*.

As stated above there is no possibility of a phase transition in this system. Any long-range magnetic order vanishes when the thermal fluctuation are comparable to the exchange energy terms, implying the system must be disordered at any $T > 0$. However, it's possible some structures to be long lived, surviving up to higher temperatures. It's under this point of view we will study the magnetic structures developed in Kekulene.

Magnetic vortices can be induced by dipole interaction as described in earlier works. When the dipolar term is sufficiently strong, the continuity of the magnetic field on the edges forces the magnetic moments to be tangent to its boundary. As a result, a magnetic vortex can be induced around the center of the nanosystem. In particular, it is important mentioning the work of Rocha et al. [30] who performed systematic Monte Carlo simulations to investigate the formation conditions of vortices in $2D$ nanodisks in several structures. It was found that there is a crossover between vortex states from capacitor-like states.

In this communication, we use Monte Carlo simulations to determine the magnetic structures present in Kekulene for several temperatures ($T$) and dipole anisotropies ($\delta = D/J$). Our results show that there are two regimes at low temperature separated by a crossover at $2.5 < \delta_{cross} < 3.0$. For $\delta < \delta_{cross}$ the ground state has a unique vortex configuration. In the region $\delta > \delta_{cross}$ arrangements of vortices-antivortices (V-AV) appears. As temperature raises the vortex structure disorders and small oscillations take over. All long the simulations the total magnetization is always zero.

## 2. Simulation Details

Our Monte Carlo simulations were done using the single flip Metropolis algorithm [31]. At low temperature we use the parallel tempering technique [32] which consists of generating $N$ replicas of the system at different temperatures. At particular intervals, in the process of equilibration, replicas $i$ and $j$ are interchanged with probability $p = \exp(\beta_i - \beta_j)(E_i - E_j)$, where $\beta = 1/k_BT$ and $E_{i(j)}$ correspond to temperature and energy of each replica respectively. This mechanism allows each system to alternate between low and high temperatures exploring a wider region in the free energy minima configuration. All quantities in our simulation are calculated in the following way. The system is thermalized using $10^4$ Monte Carlo steps (MCS). A MCS consists in an attempt to sweep all spins in the system. After the system is equilibrated a search for vortices is done. This procedure is repeated $10^3$ times. At the end, our results consists of an average over $10^3$ in equilibrium independent configurations. When not shown the error bars are smaller than the symbols. Special care was taken with the pseudo-random number generator [33, 34]. Here we used the *xoshiro*256** pseudorandom number generator [35] implemented in *gfortran-2008*. This generator has a period of $2^{256} - 1$.

To locate a vortex (antivortex) in the system we remind that a vortex (Antivortex) is a topological excitation in which spins on a closed path around the vortex core precess by $2\pi$ ($-2\pi$). The chirality $q = \pm 1$ is defined as

$$q = \frac{1}{2\pi} \sum (\Phi_i - \Phi_j) = \pm 1 , \qquad (2)$$

where, Φ is the angle between the *XY* spin vector component and some fixed in plane direction (See Fig. 1). We parameterize the spin as $\tilde{S} = |\tilde{S}|(\sin\theta \cos\varphi, \sin\theta \sin\varphi, \cos\theta)$, where $\varphi$ and $\theta$ are the spherical angles. Because $q$ depends only on the in-plane angular difference we can take Φ = $\varphi$.

## 3. Results

In Fig. 3 we show the vortex density as a function of temperature for several values of the anisotropy parameter. Here, the density is calculated as the sum of vortices that appears in any particular configuration divided by the number of independent configurations. There are two distinct regions in this picture separated by a crossover $\delta_{cross} \approx 3.0$. For higher $\delta$ > 3.0 the vortex density ($\rho$) line is always below 0.5. For $\delta$ < 3.0 the vortex density changes the slope and $\rho$ > 0.5 saturating at $\rho$ = 1 for lower $\delta$ and *T*. In Fig. 4 it is shown a complete 3*D* picture of the vortex density as a function of *T* and $\delta$. This behavior can be understood as follows. The Kekulene has two sublattices, one with 48 *C* atoms with spin 5/2 and the other with 24 hydrogen sublattice with spin 2. For high $\delta$ the vortex energy is large being excited only for large *T*. As *T* diminishes the lowest energy is for the hydrogen in the spin state *S* = 0, lowering the vortex density. For smaller values of the dipole anisotropy ($\delta$ < 3.0) the vortex energy and the hydrogen spin configuration with *S* = 0 start to compete and the vortex density grows saturating at $\rho$ = 1.

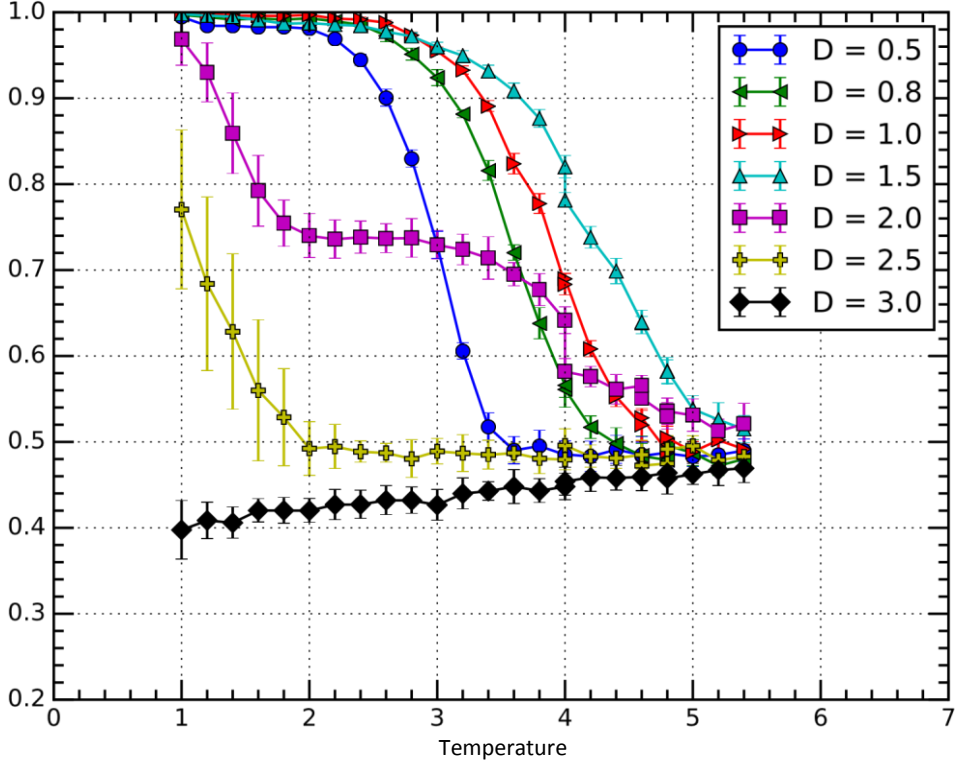

Figure 3: (Color online)Vortex density as a function of temperature for several values of $\delta = D/J$. There is a crossover for $\delta$ slightly above 3.0 where vortex density, $\rho$, changes its concavity. Below 3.0, $\rho$ < 0.5 for $\delta$ > 3.0 and low *T*, the density saturates at $\rho$ = 1. The lines are only guide to the eyes.

## 4. Final Remarks

In this work we have shown that the organic compound Kekulene can support stable vortices in a wide region of dipole anisotropy, being a real possibility for building magnetic nanostructures candidates for building advanced electronic devices with applications in several areas. Interactions between Kekulene molecules were studied by [28, 29] when the dipole interaction is absent. An important step forward is to understand how Kekulene molecules interact and its stability in the presence of dipole interactions. It should be important to experimentally test existence of vortices in Kekulene. An interesting path to explore further would be to decorate the kekulene molecule by exchanging hydrogen atoms for transition metals, which would give the system a more ferromagnetic character.

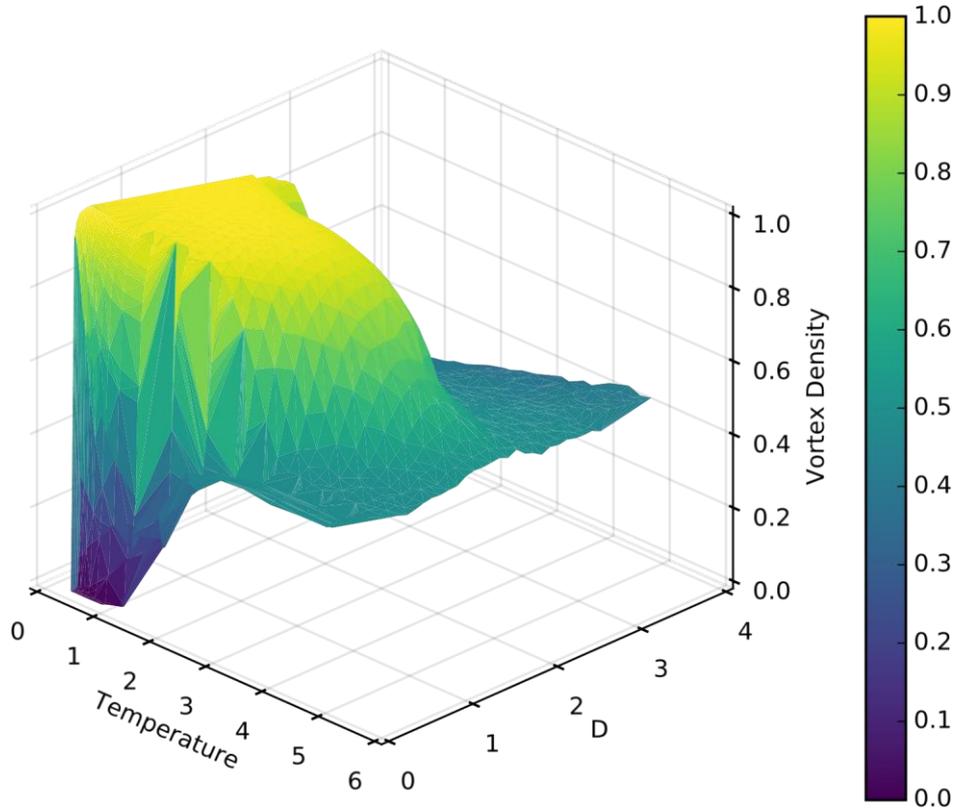

Figure 4: (Color online) 3D picture showing the density as a function of the dipole anisotropy and temperature. For low $T$ and $\delta$ the vortex density is small. There is a region for intermediate values of $\delta$ were the density saturates at $\rho = 1$ even for finite temperature. For large values of $\delta$ the density is approximately constant = 0.5. .

## Acknowledgments

This work was partially supported by CNPq and Fapemig, Brazilian Agencies. BVC thanks CNPq for the support under grant CNPq 402091/2012-4.